\documentclass[letterpaper]{article} 
\usepackage{aaai2027}  
\usepackage[hyphens]{url}  
\usepackage{graphicx} 
\usepackage{natbib}  
\usepackage{caption} 
\usepackage{algorithm}
\usepackage{algpseudocode}

\algnewcommand{\algorithmicinput}{\textbf{Input:}}
\algnewcommand{\Input}{\item[\algorithmicinput]}
\algnewcommand{\algorithmicoutput}{\textbf{Output:}}
\algnewcommand{\Output}{\item[\algorithmicoutput]}

\algrenewcommand\alglinenumber[1]{\footnotesize #1}

\usepackage{amsmath,amsfonts,amssymb}
\usepackage{booktabs}
\usepackage{multirow}
\usepackage{xspace}
\usepackage{subcaption}

\usepackage{makecell}
\usepackage{pifont}

\usepackage{amsthm}

\newtheorem{theorem}{Theorem}
\newtheorem{proposition}[theorem]{Proposition}

\newcommand{\ours}{DFCS\xspace} 
\title{Diversity Matters: Distributional Feature Coverage Sample Selection for Data-Efficient Backdoor Attacks}

\author{
Yi Yang\textsuperscript{\rm 1},
Xiaoke Chen\textsuperscript{\rm 1},
Jinyang Huang\textsuperscript{\rm 1}\corresponding,
Feng-Qi Cui\textsuperscript{\rm 2},
Yu-Tong Guo\textsuperscript{\rm 1},
Jia-Cheng Zhao\textsuperscript{\rm 1},\\
Haiming Jin\textsuperscript{\rm 3},
Xiaokang Zhou\textsuperscript{\rm 4},
Meng Li\textsuperscript{\rm 1}\corresponding
}

\affiliations{
\textsuperscript{\rm 1}
School of Computer Science and Information Engineering,
Hefei University of Technology,
Hefei 230601, China\\

\textsuperscript{\rm 2}
School of Information Science and Technology,
University of Science and Technology of China,
Hefei 230026, China\\

\textsuperscript{\rm 3}
John Hopcroft Center for Computer Science,
Shanghai Jiao Tong University,
Shanghai 200240, China\\

\textsuperscript{\rm 4}
Faculty of Business Data Science,
Kansai University,
Osaka 565-8585, Japan\\

hjy@hfut.edu.cn, mengli@hfut.edu.cn
}

\begin{document}

\maketitle

\begin{abstract}
	Backdoor attacks compromise training data so that a model retains clean
	accuracy but predicts an attacker-chosen target on triggered inputs. At very
	low poisoning rates, only a few samples convey the trigger--target
	association, making poison-sample selection critical. Existing methods
	typically rank candidates using per-sample scores, which can select redundant
	samples from similar semantic regions, and many require task-specific
	surrogate training. We propose Distributional Feature Coverage Sample
		Selection (\ours), a training-free, trigger-agnostic method that clusters
	fixed pretrained features into one region per poisoning slot and selects the
	centroid-nearest sample from each region. A local first-order analysis relates
	this allocation to feature-coverage and representative-mass terms. Across
	BadNets and Blended attacks on CIFAR-10, Tiny-ImageNet, and Imagenette, \ours
	achieves the highest mean attack success rate among seven selectors in all six
	dataset--attack settings, averaging $96.30\%$ and exceeding the strongest
	comparator in each setting by $4.60$ percentage points on average while
	preserving clean accuracy. These results support distributional feature
	coverage as an effective selection principle for low-budget dirty-label
	backdoor attacks.
\end{abstract}

\section{Introduction}

Backdoor poisoning attacks modify a small subset of the training data to
induce attacker-specified predictions on triggered inputs while preserving
standard behavior on clean inputs
~\cite{gu2017badnets,chen2017targeted}. Because larger poison sets increase
manipulation cost and exposure to filtering or inspection, recent work has
studied attacks under very low poisoning rates
~\cite{xia2022data,xun2024minimalism}. In this regime, the trigger--target
association may be conveyed by small poison sets, making the composition of the poison set as important as its
cardinality.


Existing poisoned-sample selectors estimate candidate importance from forgetting
events~\cite{xia2022data, li2023explore}, representation
distance~\cite{wu2023computation}, high-frequency
response~\cite{xun2024minimalism}, confidence~\cite{he2024stealthy}, or
influence~\cite{wei2025influence}. Most of these criteria score candidates
individually. Consequently, samples that are highly ranked in isolation may
occupy similar feature neighborhoods and provide redundant supervision when
selected together. The problem is therefore not only which candidates have
large individual scores, but how the entire budget is allocated. Pointwise
importance asks which samples are strong in isolation, and pairwise diversity
asks whether selected samples differ from one another; distributional
allocation instead asks how well the selected set represents the candidate pool
as a whole. Let $B$ denote the poisoning budget and
$\mathcal C$ the eligible pool of non-target training candidates. When
$B\ll|\mathcal C|$, assigning multiple slots to one feature
neighborhood may leave other neighborhoods unrepresented. PFS
introduces stochastic diversity after filtering by clean--poison similarity
~\cite{li2024proxy}, but it does not explicitly allocate the final budget over
the candidate feature distribution.

We address this gap through a backdoor-specific formulation:
low-budget dirty-label selection is treated as allocation of $B$
equal-weight poisoning slots over the clean candidate distribution. We
evaluate distributional feature coverage as the resulting empirical principle
and formulate it as Distributional Feature Coverage Sample
Selection (\ours). DFCS uses a frozen external encoder, $k$-means with $B$
regions, and centroid-nearest projection to real candidates. The budget thus determines both the number of represented regions and
the number of selected samples, while each region contributes one actual
training example. This coupling discourages redundant selection and distributes
the scarce slots across complementary feature regions. 


We further provide a local first-order motivation
tailored to dirty-label replacement. Under the equal weighting used by the implemented attack, the
discrepancy between the candidate-average and selected-set replacement
signals is bounded in terms of feature-space quantization and
representative-mass imbalance. This decomposition separates error caused by
incomplete feature coverage from distortion caused by assigning equal
training weight to regions with unequal candidate mass. It therefore provides
a theoretical rationale for coverage-based selection and yields measurable
component-wise predictions. For a fixed $k$-means partition, we additionally
show that the centroid-nearest real candidate is the exact discrete minimizer
of within-region squared distortion. We test the local predictions using shared-checkpoint
component diagnostics and matched end-to-end controls. The diagnostics separate
feature coverage from region-mass mismatch, while histogram-matched,
class-stratified, and cluster-random comparisons probe class composition and
representative choice.

In summary, this paper makes the following contributions:
\begin{itemize}
\item We formulate low-budget dirty-label selection as equal-weight
allocation over complementary candidate regions and instantiate it as
trigger-independent, training-free DFCS using established clustering machinery.
\item We further motivate the design of DFCS through a local first-order analysis that decomposes dirty-label replacement-signal approximation error into feature-coverage and representative-mass terms.
\item We show that DFCS achieves strong attack success across
CIFAR-10, Tiny-ImageNet, and Imagenette under low poisoning budgets,
while remaining effective across attacks, victim architectures, pretrained
encoders, and partitioning objectives.
\end{itemize}

\section{Related Work}

\paragraph{Poisoning-based backdoor attacks.}
Backdoor attacks cause a compromised model to behave normally on clean
inputs while producing an attacker-specified prediction when a trigger
is present. We focus on the poison-only setting, where the attacker
can manipulate part of the training data but cannot control the
victim's architecture or training procedure. Representative trigger
constructions include patch-based BadNets~\cite{gu2017badnets}, image
blending~\cite{chen2017targeted}, and adaptive attacks~\cite{qi2023revisiting}. For a broad range of poisoning-based backdoor attacks, readers are referred to existing surveys~\cite{li2022backdoor,wu2025backdoorbench}.
These studies primarily alter trigger or injection mechanisms, whereas our
work addresses which samples to poison under a limited dirty-label budget.

\paragraph{Poisoned-sample selection.}
Prior work exploits unequal sample contributions to backdoor
implantation. FUS~\cite{xia2022data} uses forgetting events, RD
~\cite{wu2023computation} measures prediction--target distance, and
other methods, select
low-confidence samples~\cite{he2024stealthy}, or estimate influence
on backdoor risk~\cite{wei2025influence}. Although their criteria
differ, these methods require task-specific training, gradients,
training dynamics, or influence estimation. Alternatives without
task-specific selector training
include HFE~\cite{xun2024minimalism}, which scores samples by
high-frequency energy, and PFS~\cite{li2024proxy}, which filters
candidates by clean--poison feature similarity and then samples from
the retained pool to encourage diversity. Neither explicitly optimizes
the distributional coverage of the final poison set. Generic subset-selection
tools include average-distortion coresets, medoid cost, worst-case $k$-center
coverage, and facility location~\cite{feldman2020coresets,sener2018active,
wei2015submodularity}. These objectives provide useful machinery for
representing a data distribution, but do not by themselves specify what should
be represented when each selected example is trigger-transformed, relabeled,
and inserted once into victim training. Our formulation makes this equal-weight
dirty-label constraint explicit: the $B$ poisoning slots are allocated over the
frozen candidate distribution, and the analysis separates within-region
coverage from mismatch between region mass and equal poison weight. We use
$k$-means as the allocation mechanism and evaluate $k$-medoids and $k$-center
as controls for distinct coverage objectives. Accordingly, our contribution is not a new generic clustering
objective, but a backdoor-specific interpretation of coverage: the $B$
equal-weight poisoning slots are allocated over the clean candidate
distribution before the selected examples are transformed and relabeled, and
this mapping is evaluated empirically.

Clean-label backdoor selection addresses a different optimization problem:
labels are preserved, and candidates may be restricted to the target class
~\cite{turner2019clean,GAO2023109512,nguyen2025wicked}. Its criteria need
to satisfy the clean-label constraint, whereas our dirty-label adversary selects
non-target samples and relabels them as the target. Clean-label methods are
therefore complementary rather than directly comparable baselines for the
threat model studied in our work.

\begin{figure}[t]
	\centering
	\includegraphics[width=\linewidth]{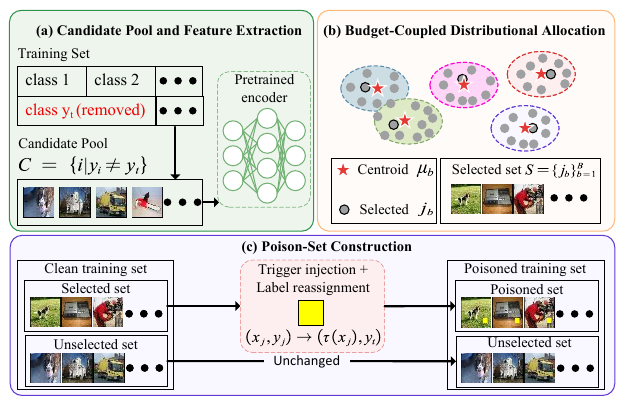}
	\caption{\textbf{Overview of DFCS.}
		(a) Non-target samples form the candidate pool and are encoded into
		normalized pretrained features.
		(b) DFCS partitions the features into $B$ clusters and selects the
		sample nearest each centroid.
		(c) Selected samples are trigger-transformed and relabeled, while
		unselected samples remain clean.}
	\label{fig:method_overview}
\end{figure}

\section{Threat Model}

\paragraph{Adversary knowledge.}
We consider a black-box, poison-only backdoor attack. The adversary has access
to the released clean training set
$\mathcal D=\{(x_i,y_i)\}_{i=1}^{N}$ and specifies a target label $y_t$ and
trigger transformation $\tau(\cdot)$. It may use a publicly released,
externally pretrained encoder $\phi$ and its documented preprocessing, but
does not train or adapt this encoder on the victim task. It has no knowledge
of the victim's architecture, parameters, optimizer, initialization, or
training procedure.

\paragraph{Adversary capability.}
The adversary may alter only the released training data and cannot influence
victim training. Under poisoning rate $r$, its budget is
$B=\lfloor rN\rfloor$. It selects at most $B$
non-target samples indexed by $\mathcal S$, applies the trigger, and relabels
them as $y_t$, yielding the poisoned training set
\[
\widetilde{\mathcal{D}}
=
\left(
\mathcal{D}
\setminus
\{(x_j,y_j)\mid j\in\mathcal{S}\}
\right)
\cup
\{(\tau(x_j),y_t)\mid j\in\mathcal{S}\}.
\]
The victim trains on $\widetilde{\mathcal D}$ using its own pipeline. The
attack succeeds when triggered non-target inputs are mapped to $y_t$ while
clean-input performance remains close to that of normal training.
This relabeling capability defines the dirty-label setting considered
throughout this work. The frozen external encoder is used only to choose
indices before poisoning.


\section{Methodology}

Under a very low poisoning budget, independently ranking candidates can
allocate several scarce slots to similar feature regions. We instead treat
the poison budget as a set-level allocation constraint: each selected sample
should represent a complementary part of the eligible candidate
distribution.

DFCS implements this backdoor-specific principle through budget-coupled
distributional allocation. It constructs the non-target candidate pool,
encodes it with a frozen pretrained model, uses $k$-means to partition the
features into exactly $B$ regions, and selects one real representative per
region. Thus, the attack allocates its $B$ equal-weight training examples over
$B$ complementary regions before poisoning the selected
samples. Figure~\ref{fig:method_overview} illustrates these three stages.

\subsection{Candidate Pool and Feature Extraction}

Following the threat model, we first construct the candidate index set containing all
non-target training samples: $\mathcal{C}=\{i\mid y_i\neq y_t\}$. Samples belonging to the target class are excluded only from selection
and remain unchanged in the training set.

We then use a fixed pretrained encoder
$\phi:\mathcal{X}\rightarrow\mathbb{R}^{d}$ to extract the
features of the candidate pool. For each
$i\in\mathcal{C}$, its normalized feature is $h_i
=
\phi(x_i)/
\|\phi(x_i)\|_2$. The encoder remains frozen throughout selection. Feature normalization
removes differences in representation magnitude and makes the
subsequent distance computation depend on direction in the representation
space. This stage requires neither victim-model information nor
task-specific surrogate training.

\subsection{Budget-Coupled Distributional Allocation}

Given the candidate pool $\mathcal{C}$ and poisoning budget $B$, let
$\mathcal{S}\subseteq\mathcal{C}$ denote a selected index set with
$|\mathcal{S}|=B$. We quantify how well $\mathcal{S}$ represents the candidate
distribution using the average feature-coverage error
\begin{equation}
	\Delta_2(\mathcal{S})
	=
	\frac{1}{|\mathcal{C}|}
	\sum_{i\in\mathcal{C}}
	\min_{s\in\mathcal{S}}
	\|h_i-h_s\|_2^2.
	\label{eq:coverage_error}
\end{equation}
A small $\Delta_2(\mathcal{S})$ indicates that candidates are, on average,
close to a selected representative. This set-level objective favors
complementary representatives that jointly cover the candidate distribution,
rather than selecting samples independently or promoting pairwise diversity
without regard to the remaining candidates.

Directly optimizing Eq.~\eqref{eq:coverage_error} over all size-$B$ subsets is
combinatorial. We therefore first optimize a tractable continuous-centroid
surrogate using $k$-means. Because the implemented poison set contains exactly
$B$ examples, each inserted once into victim training, we set the number of
clusters equal to the poisoning budget:
\begin{equation}
	\{\mu_b\}_{b=1}^{B}
	=
	\arg\min_{\{\mu_b\}_{b=1}^{B}}
	\sum_{i\in\mathcal{C}}
	\min_{b\in\{1,\ldots,B\}}
	\|h_i-\mu_b\|_2^2,
	\label{eq:dfcs_kmeans}
\end{equation}
where $\mu_b$ is the centroid of the induced cluster $\mathcal{C}_b$. This
budget coupling allocates one poisoning slot to each of the $B$ feature
regions discovered in the candidate distribution.

Because the centroids are generally not training samples, they cannot be
selected directly. We therefore project each centroid onto a real candidate
from its corresponding cluster:
\begin{equation}
	j_b
	=
	\arg\min_{i\in\mathcal{C}_b}
	\|h_i-\mu_b\|_2^2,
	\qquad
	b=1,\ldots,B.
	\label{eq:nearest_representative}
\end{equation}
The resulting poison-selection index set is
$\mathcal{S}=\{j_b\}_{b=1}^{B}$.

For a fixed learned partition, centroid-nearest projection also minimizes the
within-cluster squared distortion among real candidates. Specifically, for
any $s\in\mathcal{C}_b$, the centroid identity gives
\begin{equation}
	\sum_{i\in\mathcal{C}_b}\|h_i-h_s\|_2^2
	=
	\sum_{i\in\mathcal{C}_b}\|h_i-\mu_b\|_2^2
	+
	|\mathcal{C}_b|\|h_s-\mu_b\|_2^2.
	\label{eq:fixed_partition_identity_main}
\end{equation}
Because the first term is independent of $s$,
Eq.~\eqref{eq:nearest_representative} selects the exact discrete minimizer
within each learned region. Appendix provides the full
derivation and relates the resulting quantization error to
$\Delta_2(\mathcal{S})$.

DFCS separates distributional allocation from realizable-sample
selection: $k$-means allocates the $B$ poisoning slots across the candidate
distribution, and centroid-nearest projection assigns one valid training
example to each slot.

\begin{algorithm}[t]
	\small
	\caption{DFCS}
	\label{alg:dfcs}
	\begin{algorithmic}[1]
		
		\Input Training set $\mathcal{D}$, target label $y_t$, poisoning
		budget $B$, and fixed pretrained encoder $\phi$
		\Output Selected index set $\mathcal{S}$
		
		\State $\mathcal{C}\gets
		\{i\in\{1,\ldots,N\}\mid y_i\neq y_t\}$
		\Comment{Candidate pool}
		
		\ForAll{$i\in\mathcal{C}$}
		\State $h_i\gets
		\phi(x_i)/\|\phi(x_i)\|_2$
		\Comment{Extracted feature}
		\EndFor
		
		\State $\{\mathcal{C}_b,\mu_b\}_{b=1}^{B}
		\gets
		\operatorname{KMeans}
		\bigl(\{h_i\}_{i\in\mathcal{C}},B\bigr)$
		
		\State $\mathcal{S}\gets\emptyset$
		
		\For{$b=1$ to $B$}
		\State $j_b\gets
		\displaystyle
		\arg\min_{i\in\mathcal{C}_b}
		\|h_i-\mu_b\|_2^2$
		\Comment{Sample selection}
		\State $\mathcal{S}\gets\mathcal{S}\cup\{j_b\}$
		\EndFor
		
		\State \textbf{return} $\mathcal{S}$
		
	\end{algorithmic}
\end{algorithm}

\subsection{Poison-Set Construction}

DFCS outputs only the selected index set $\mathcal{S}$ and does not
access the trigger pattern or trigger-injection function during
selection. After selection, a compatible dirty-label backdoor attack can apply its
own poisoning transformation to the samples indexed by $\mathcal{S}$.
Under the setting considered in our threat model, the resulting poisoned subset is
$\mathcal{P}
=
\left\{
\bigl(\tau(x_s),y_t\bigr)
\;\middle|\;
s\in\mathcal{S}
\right\}$.
The samples not indexed by $\mathcal{S}$ remain unchanged, and
$\mathcal{P}$ replaces the corresponding selected clean samples
according to the poisoned-training-set construction defined in the
\emph{Threat Model} section.

This construction is external to DFCS. In particular, the selected set
$\mathcal{S}$ depends only on the clean candidate distribution, the
pretrained encoder, and the poisoning budget. It is therefore
independent of the trigger pattern, injection mechanism, victim
architecture, and victim-training configuration. For a fixed dataset,
target label, budget, and encoder, the same indices can consequently be
paired with different trigger transformations without rerunning selection.
Algorithm~\ref{alg:dfcs} summarizes this trigger-independent-at-selection
process.

\paragraph{Computational complexity.}
Let $n=|\mathcal C|$, $d$ be the feature dimension, and $C_\phi$
denote the cost of one encoder forward pass. With at most $T$ $k$-means
iterations, DFCS requires
$O(nC_\phi+TnBd)$ time and $O(nd+Bd)$ memory. Projecting the centroids
onto real candidates adds $O(nd)$ time, which is dominated by clustering
for typical $B$ and $T$. Thus, for a fixed encoder, poisoning budget, and
clustering configuration, selection scales linearly with the candidate-pool
size and requires no victim- or task-specific surrogate training.

\subsection{Local Equal-Weight Signal Approximation}

DFCS's poison-set construction replaces each selected clean example once.
Consequently, within the poison subset, every representative in
$\mathcal S$ has the same weight $1/B$. To motivate the coverage criterion
used by DFCS, we analyze how this equal-weight selected set approximates the
replacement signal obtained by averaging over the full candidate pool. The
analysis is local to reference parameters $\theta_0$ and isolates two
measurable approximation terms: feature-coverage error and
representative-mass mismatch. Let $n=|\mathcal C|$, $B=|\mathcal S|$, and
let $\Delta_2(\mathcal S)$ be the coverage error defined in
Eq.~\eqref{eq:coverage_error}.

For an evaluation input $x$, let
$q_x=\nabla_\theta r_{\theta_0}(x)$ denote the gradient of a smooth triggered
target margin, whose definition is given in Appendix.
Consistent with this replacement operation, we define the gradient change
associated with candidate $i$ and its projected local effect as
\begin{equation}
	\begin{aligned}
		d_i
		&=
		\nabla_\theta\ell(f_{\theta_0}(\tau(x_i)),y_t)
		-
		\nabla_\theta\ell(f_{\theta_0}(x_i),y_i),
		\\[-1mm]
		A_x(i)
		&=
		-q_x^\top d_i.
	\end{aligned}
	\label{eq:gradient_terms}
\end{equation}
Thus, $d_i$ captures the gradient change induced by replacing the clean
example $(x_i,y_i)$ with its triggered, target-labeled counterpart.

Let $\eta>0$ denote a common first-order scale that absorbs the factor
$B/N$. Averaging the local effects over the candidate pool gives the
candidate-reference signal
\begin{equation}
	\widetilde{\Delta}_{\mathcal C}(x)
	=
	\eta\frac{1}{n}\sum_{i\in\mathcal C}A_x(i),
	\label{eq:full_candidate_signal}
\end{equation}
whereas the implemented equal-weight poison set gives
\begin{equation}
	\widetilde{\Delta}_{\mathcal S}^{u}(x)
	=
	\eta\frac{1}{B}\sum_{s\in\mathcal S}A_x(s).
	\label{eq:equal_selected_signal_main}
\end{equation}

Let $\pi_{\mathcal S}(i)$ denote the nearest selected representative of
candidate $i$, and let
\[
w_s
=
\frac{1}{n}
\left|
\{i\in\mathcal C:\pi_{\mathcal S}(i)=s\}
\right|
\]
be the fraction of the candidate pool represented by
$s\in\mathcal S$. The candidate regions can have unequal masses, whereas the
implemented poison set assigns every representative weight $1/B$. We measure
the resulting representative-mass mismatch by
\begin{equation}
	\Gamma(\mathcal S)
	=
	\sum_{s\in\mathcal S}
	\left|w_s-\frac{1}{B}\right|.
	\label{eq:cluster_imbalance_main}
\end{equation}

\begin{proposition}[Equal-weight local signal approximation]
	\label{prop:coverage_main}
	Assume that $A_x$ is $L_A(x)$-Lipschitz on the candidate pool with respect
	to the normalized pretrained features and that
	$|A_x(s)|\le M_A(x)$ for every $s\in\mathcal S$. Then
	\begin{equation}
		\left|
		\widetilde{\Delta}_{\mathcal C}(x)
		-
		\widetilde{\Delta}_{\mathcal S}^{u}(x)
		\right|
		\le
		\eta L_A(x)\sqrt{\Delta_2(\mathcal S)}
		+
		\eta M_A(x)\Gamma(\mathcal S).
		\label{eq:equal_weight_bound_main}
	\end{equation}
\end{proposition}

Proposition~\ref{prop:coverage_main} separates the approximation error into
two components. The first depends on $\Delta_2(\mathcal S)$ and therefore
directly reflects the distributional-coverage objective in
Eq.~\eqref{eq:coverage_error}. The second captures the discrepancy
between the candidate mass represented by each selected sample and the
uniform weight imposed by the poison-set construction.

For the $k$-means partition and centroid-nearest representatives used by
DFCS, Appendix proves
\[
\Delta_2(\mathcal S)
\le
J_{\mathrm{rep}}
\le
2J_{\mathrm{KM}},
\]
where $J_{\mathrm{KM}}$ and $J_{\mathrm{rep}}$ denote the normalized
centroid and projected-representative distortions, respectively. Thus, the
allocation rule targets an upper bound on the coverage component, while
$\Gamma(\mathcal S)$ provides a complementary diagnostic of equal-weight
region-mass mismatch. The mechanism diagnostics and controlled comparisons
below evaluate these two components using coverage, mass-imbalance,
signal-error, and controlled-comparison diagnostics.

\section{Experiments}

\begin{table*}[t]
	\centering
	
		\begin{tabular}{lc
				r@{\,\ensuremath{\pm}\,}l
				r@{\,\ensuremath{\pm}\,}l |
				r@{\,\ensuremath{\pm}\,}l
				r@{\,\ensuremath{\pm}\,}l |
				r@{\,\ensuremath{\pm}\,}l
				r@{\,\ensuremath{\pm}\,}l}
			\toprule
			\multirow{2}{*}{Attack}
			& \multirow{2}{*}{Selector}
			& \multicolumn{4}{c|}{CIFAR-10}
			& \multicolumn{4}{c|}{Tiny-ImageNet}
			& \multicolumn{4}{c}{Imagenette} \\
			\cmidrule(lr){3-6}
			\cmidrule(lr){7-10}
			\cmidrule(lr){11-14}
			& & \multicolumn{2}{c}{ACC \ensuremath{\uparrow}}
			& \multicolumn{2}{c|}{ASR \ensuremath{\uparrow}}
			& \multicolumn{2}{c}{ACC \ensuremath{\uparrow}}
			& \multicolumn{2}{c|}{ASR \ensuremath{\uparrow}}
			& \multicolumn{2}{c}{ACC \ensuremath{\uparrow}}
			& \multicolumn{2}{c}{ASR \ensuremath{\uparrow}} \\
			\midrule
			
			No Attack & --
			& 94.38 & 0.13 & \multicolumn{2}{c|}{--}
			& 64.45 & 0.31 & \multicolumn{2}{c|}{--}
			& 82.03 & 0.98 & \multicolumn{2}{c}{--} \\
			
			\midrule
			
			\multirow{7}{*}{BadNets}
			& RS
			& 94.26 & 0.31 & 81.22 & 2.28
			& 64.48 & 0.27 & 83.79 & 2.33
			& 82.49 & 0.99 & 68.34 & 14.71 \\
			
			& FUS
			& 94.26 & 0.10 & 86.07 & 3.85
			& 64.58 & 0.20 & 86.02 & 2.38
			& 82.80 & 1.22 & 78.68 & 5.70 \\
			
			& RD
			& 94.16 & 0.30 & 91.06 & 4.09
			& 64.60 & 0.27 & 89.48 & 0.33
			& 82.51 & 1.27 & 87.39 & 3.21 \\
			
			& PFS
			& 94.23 & 0.18 & 85.69 & 4.44
			& 64.31 & 0.27 & 83.37 & 1.66
			& 82.97 & 0.88 & 83.35 & 5.02 \\
			
			& HFE
			& 94.30 & 0.20 & 84.32 & 3.94
			& 64.32 & 0.15 & 86.78 & 2.46
			& 83.01 & 1.36 & 84.51 & 3.75 \\
			
			& IFS
			& 94.33 & 0.18 & 86.15 & 5.48
			& 64.39 & 0.30 & 91.06 & 1.45
			& 82.70 & 0.66 & 85.35 & 4.39 \\
			
			& DFCS (ours)
			& 94.53 & 0.33 & \textbf{98.32} & \textbf{0.74}
			& 64.57 & 0.29 & \textbf{93.81} & \textbf{0.57}
			& 81.96 & 1.01 & \textbf{96.10} & \textbf{1.68} \\
			
			\midrule
			
			\multirow{7}{*}{Blended}
			& RS
			& 94.28 & 0.28 & 87.30 & 4.39
			& 64.98 & 0.17 & 93.25 & 0.38
			& 82.87 & 0.17 & 86.31 & 3.87 \\
			
			& FUS
			& 94.42 & 0.24 & 88.40 & 3.38
			& 64.80 & 0.25 & 93.80 & 1.77
			& 82.79 & 0.47 & 92.59 & 3.19 \\
			
			& RD
			& 94.32 & 0.16 & 92.76 & 1.38
			& 64.62 & 0.26 & 94.41 & 0.56
			& 83.04 & 0.97 & 91.72 & 3.38 \\
			
			& PFS
			& 94.29 & 0.13 & 93.66 & 0.57
			& 64.44 & 0.40 & 93.94 & 1.95
			& 82.75 & 0.20 & 89.48 & 2.77 \\
			
			& HFE
			& 94.36 & 0.27 & 88.50 & 2.43
			& 64.90 & 0.15 & 93.89 & 2.70
			& 82.80 & 0.74 & 89.05 & 2.67 \\
			
			& IFS
			& 94.40 & 0.19 & 89.37 & 1.36
			& 64.42 & 0.10 & 94.15 & 1.66
			& 83.11 & 1.09 & 88.71 & 4.03 \\
			
			& DFCS (ours)
			& 94.32 & 0.21 & \textbf{96.40} & \textbf{1.04}
			& 64.61 & 0.69 & \textbf{98.19} & \textbf{1.05}
			& 82.77 & 1.14 & \textbf{94.97} & \textbf{2.00} \\
			
			\bottomrule
		\end{tabular}%
	\caption{
		Five-run results of poisoned-sample selection, reported as mean
		\ensuremath{\pm} sample standard deviation. 
	}
	\label{tab:main_acc_asr}
\end{table*}

\subsection{Experimental Settings}

\paragraph{Datasets and attacks.}
We evaluate DFCS on CIFAR-10~\cite{krizhevsky2009learning},
Tiny-ImageNet~\cite{le2015tiny}, and
Imagenette~\cite{howard2019imagenette,deng2009imagenet}. BadNets
~\cite{gu2017badnets} and Blended~\cite{chen2017targeted} serve as the
main evaluation attacks because they are widely used in data-efficient
backdoor and poisoned-sample selection studies
~\cite{xia2022data,wu2023computation,xun2024minimalism,
	li2024proxy,wei2025influence}. They also represent two distinct trigger
constructions under our threat model: a localized patch and an image-wide
blended pattern.

We use a poisoning rate of $0.04\%$ for both attacks on CIFAR-10
($B=20$). On Tiny-ImageNet, the rates are $0.15\%$ for BadNets
($B=150$) and $0.10\%$ for Blended ($B=100$). On Imagenette, the
corresponding rates are $0.25\%$ ($B=23$) and $1.00\%$ ($B=94$).
These settings focus the evaluation on very low poisoning budgets, where
the allocation of individual poisoning slots is consequential. Complete
attack configurations are provided in Appendix;
Tables~\ref{tab:dataset_statistics} and~\ref{tab:poison_counts} report the
dataset statistics and integer payload budgets. Additional attack evaluations
are reported separately in Appendix.

\paragraph{Baselines.}
We compare DFCS with random selection (RS), FUS
~\cite{xia2022data}, RD~\cite{wu2023computation}, PFS
~\cite{li2024proxy}, HFE~\cite{xun2024minimalism}, and IFS
~\cite{wei2025influence}. These baselines span uninformed sampling,
training-dynamics criteria, model-response distance, feature-based filtering,
input-frequency statistics, and influence-based selection. Their
implementation, configuration, and provenance details are provided in
Appendix.

\paragraph{Models and evaluation protocol.}
Unless otherwise stated, class 0 is the target class and
ResNet-18~\cite{he2016deep} is the victim architecture. DFCS and PFS use
the same frozen DINOv3 ViT-S/16 encoder~\cite{dinov3}, while baselines
requiring a task-specific surrogate use
PreActResNet-18~\cite{he2016identity}. Within each dataset--attack
setting, all selectors receive the same eligible non-target candidate pool,
target class, trigger implementation, integer budget, and victim-training
configuration. This common protocol isolates the effect of the
poisoned-sample selection rule. Complete selector, encoder, victim-training,
and hardware configurations are reported in Appendix.

We report the mean and sample standard deviation over five
selection-and-training runs for each setting. The runs are matched across
selectors using the same master seeds within each dataset--attack setting.
Means, sample standard deviations, and paired differences are interpreted
as descriptive summaries. The complete run and seed protocol is provided in
Appendix.

\paragraph{Evaluation metrics.}
We report clean accuracy (ACC) on unmodified test samples and attack success
rate (ASR), defined as the fraction of triggered non-target test samples
classified as the target class. Their formal definitions are given in
Appendix. Selection time includes all
preprocessing, feature or score computation, proxy training, and subset
construction required to produce the selected indices, while excluding
victim training.

\subsection{Main Results}

\paragraph{Attack effectiveness and clean accuracy.}
Table~\ref{tab:main_acc_asr} compares DFCS with six baseline selectors.
DFCS achieves the highest mean ASR in all six dataset--attack settings,
ranging from $93.81\%$ to $98.32\%$, and is the only selector whose mean
ASR exceeds $90\%$ in every setting. Its margin over the best competing
baseline in each setting ranges from $2.38\%$ to $8.71\%$.
Averaged across the three datasets, this margin is $6.24\%$
for BadNets and $2.97\%$ for Blended. The difference between
DFCS's mean ACC and that of the corresponding no-attack model ranges from
$-0.07\%$ to $+0.74\%$ across the six settings.

We additionally evaluate the $B=20$ CIFAR-10/BadNets setting under fixed
Neural Cleanse~\cite{wang2019neural}, Beatrix~\cite{ma2022beatrix}, and
MM-BD~\cite{wang2024mmbd} defenses; details are reported in
Appendix.

\paragraph{Selection efficiency.}
Table~\ref{tab:selection_time} reports the end-to-end time required to
produce the selected poison indices. HFE is the fastest non-random selector
on CIFAR-10 and Tiny-ImageNet, requiring 2 and 8 seconds, respectively,
whereas DFCS is slightly faster on Imagenette, requiring 5 seconds compared
with HFE's 6 seconds. Across the three datasets, DFCS is faster than FUS, RD, PFS, and IFS in every case.
Averaging across the six dataset--attack settings, DFCS achieves
$96.30\%$ mean ASR, compared with $87.84\%$ for HFE.

\begin{table}[t]
	\centering
	\small

	\begin{tabular}{lccc}
		\toprule
		Selection method & CIFAR-10 & Tiny-ImageNet & Imagenette \\
		\midrule
		RS    & $<1$        & $<1$         & $<1$       \\
		FUS   & $3{,}542$   & $18{,}901$   & $2{,}800$ \\
		RD    & $354$       & $795$        & $302$     \\
		PFS   & $43$        & $101$        & $12$      \\
		HFE   & $2$         & $8$          & $6$       \\
		IFS   & $2{,}828$   & $21{,}329$   & $671$     \\
		\ours & $21$        & $47$         & $5$       \\
		\bottomrule
	\end{tabular}
	\caption{Mean end-to-end sample-selection time in seconds, averaged over
		BadNets and Blended runs for each dataset. Victim training
		is excluded for all methods.}
	\label{tab:selection_time}
\end{table}

\begin{figure*}[t]
    \centering
    \newcommand{\toprowheight}{3.8cm}
    \newcommand{\bottomrowheight}{3.8cm}

    \setcounter{subfigure}{0}
    \begin{subfigure}[b]{0.32\textwidth}
        \centering
        \begin{minipage}[b][\toprowheight][b]{\linewidth}
            \centering
            \includegraphics[
                width=\linewidth,
                height=\toprowheight,
                keepaspectratio
            ]{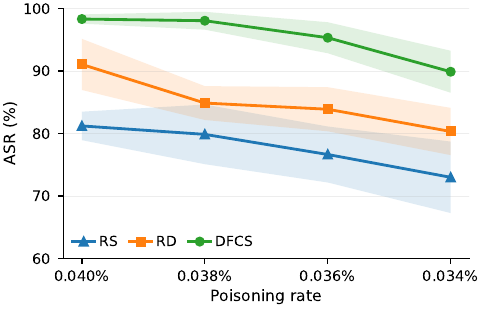}
        \end{minipage}
        \caption{Poisoning rate}
        \label{fig:result-a}
    \end{subfigure}
    \hfill
    \setcounter{subfigure}{2}
    \begin{subfigure}[b]{0.32\textwidth}
        \centering
        \begin{minipage}[b][\toprowheight][b]{\linewidth}
            \centering
            \includegraphics[
                width=\linewidth,
                height=\toprowheight,
                keepaspectratio
            ]{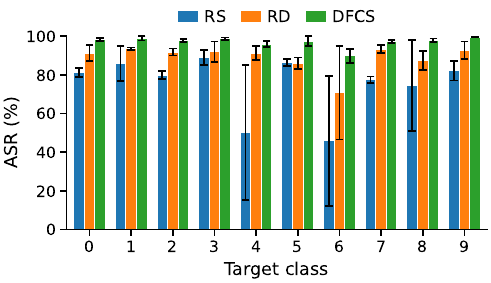}
        \end{minipage}
        \caption{Target class}
        \label{fig:result-c}
    \end{subfigure}
    \hfill
    \setcounter{subfigure}{4}
    \begin{subfigure}[b]{0.32\textwidth}
        \centering
        \begin{minipage}[b][\toprowheight][b]{\linewidth}
            \centering
            \includegraphics[
                width=\linewidth,
                height=\toprowheight,
                keepaspectratio
            ]{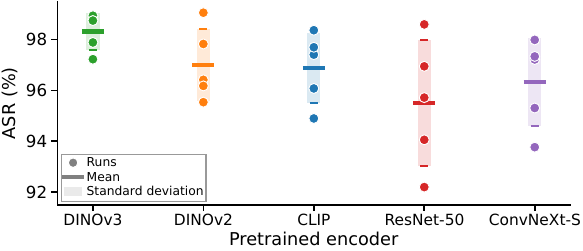}
        \end{minipage}
        \caption{Pretrained encoder}
        \label{fig:result-e}
    \end{subfigure}\par

    \setcounter{subfigure}{1}
    \begin{subfigure}[b]{0.32\textwidth}
        \centering
        \begin{minipage}[b][\bottomrowheight][b]{\linewidth}
            \centering
            \includegraphics[
                width=\linewidth,
                height=\bottomrowheight,
                keepaspectratio
            ]{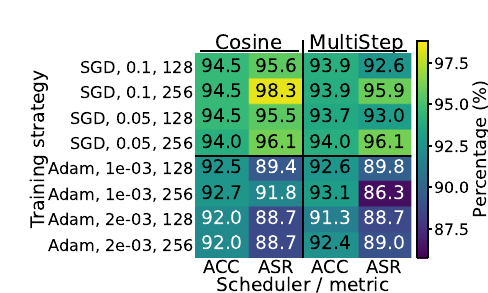}
        \end{minipage}
        \caption{Victim-training strategy}
        \label{fig:result-b}
    \end{subfigure}
    \hfill
    \setcounter{subfigure}{3}
    \begin{subfigure}[b]{0.32\textwidth}
        \centering
        \begin{minipage}[b][\bottomrowheight][b]{\linewidth}
            \centering
            \includegraphics[
                width=\linewidth,
                height=\bottomrowheight,
                keepaspectratio
            ]{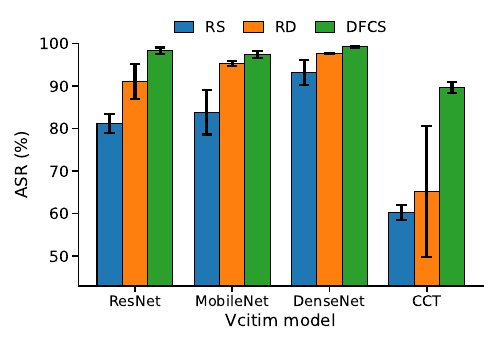}
        \end{minipage}
        \caption{Victim architecture}
        \label{fig:result-d}
    \end{subfigure}
    \hfill
    \setcounter{subfigure}{5}
    \begin{subfigure}[b]{0.32\textwidth}
        \centering
        \begin{minipage}[b][\bottomrowheight][b]{\linewidth}
            \centering
            \includegraphics[
                width=\linewidth,
                height=\bottomrowheight,
                keepaspectratio
            ]{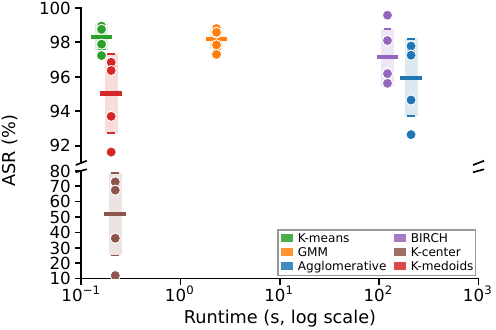}
        \end{minipage}
        \caption{Allocation methods}
        \label{fig:result-f}
    \end{subfigure}

    \caption{\textbf{Ablation and robustness studies with BadNets on CIFAR-10.}
    	We vary (a) the local poison budget, (b) victim-training strategy,
    	(c) target class, (d) victim architecture, (e) frozen pretrained encoder,
    	and (f) allocation method. Reported ACC
    	and ASR values are five-run means; where shown, error bars or shaded
    	intervals denote sample standard deviations, and panels (e) and (f)
    	additionally show individual runs. Panels (b)--(f) use the default budget
    	$B=20$. Runtime in panel (f) covers the allocation stage only and excludes
    	feature extraction, whereas Table~\ref{tab:selection_time} reports
    	end-to-end selection time.}
    \label{fig:six-subfigures}
\end{figure*}

\subsection{Ablation and Robustness Studies}

Figure~\ref{fig:six-subfigures} presents six ablation and robustness
evaluations for BadNets/CIFAR-10. Unless otherwise stated, all experiments
use the common configuration described under \emph{Experimental Settings}.
Where included, RS provides an uninformed reference, while RD represents the
strongest baseline on average in Table~\ref{tab:main_acc_asr}.
Appendix reports the corresponding evaluations
for CIFAR-10/Blended and for both attacks on Tiny-ImageNet and Imagenette.

\subsubsection{Local Budget Sensitivity}
Figure~\ref{fig:result-a} compares DFCS with RS and RD over a local range
around the default poisoning budget. The sweep uses
$B\in\{17,18,19,20\}$, corresponding to poisoning rates of
$0.034\%$, $0.036\%$, $0.038\%$, and $0.040\%$, respectively.
DFCS maintains a higher mean ASR than both references throughout this
very-low-budget range.

\subsubsection{Victim-Training Strategy}
Figure~\ref{fig:result-b} varies the optimizer, learning rate, batch size,
and learning-rate scheduler. Across all configurations, DFCS
maintains a mean ASR of at least $86.3\%$, indicating that its effectiveness
does not depend on a single victim-training strategy. Under SGD, DFCS achieves
$92.6\%$--$98.3\%$ ASR with $93.7\%$--$94.5\%$ ACC, while the evaluated
Adam configurations yield $86.3\%$--$91.8\%$ ASR with
$91.3\%$--$93.1\%$ ACC. These results demonstrate robustness across the
evaluated optimizers, learning rates, batch sizes, and schedulers, while also
showing that the attack strength is higher under SGD than under Adam in this attack setting.

\subsubsection{Target Class}
Figure~\ref{fig:result-c} evaluates all ten CIFAR-10 classes as attack
targets. DFCS achieves the highest mean ASR for every target, attaining
approximately $95\%$--$99\%$ for nine classes and about $89\%$ for
class 6. For class 6, RD and RS decrease to approximately
$70\%$ and $45\%$, respectively, with greater dispersion. This shows that DFCS exhibits less target-dependent variation than
the two reference selectors.

\subsubsection{Victim Architecture}
Figure~\ref{fig:result-d} evaluates three convolutional architectures:
ResNet-18, MobileNetV3-Small~\cite{howard2019searching}, and
DenseNet-121~\cite{huang2017densely}, together with the Compact
Convolutional Transformer (CCT)~\cite{hassani2021escaping}, a
transformer-based victim model. DFCS achieves approximately $98.3\%$,
$97.4\%$, $99.2\%$, and $89.7\%$ ASR on these architectures,
respectively, exceeding RD in every case. These results show that DFCS remains
effective across both convolutional and transformer-based victims. Although
its ASR is lower on CCT than on the three CNNs, its advantage over RD is
preserved, indicating that the selected poison set transfers across distinct
victim-model families.

\subsubsection{Pretrained Encoder}
Figure~\ref{fig:result-e} compares DINOv3,
DINOv2~\cite{oquab2024dinov2}, CLIP~\cite{radford2021learning},
ResNet-50~\cite{he2016deep}, and
ConvNeXt-S~\cite{liu2022convnet}. Their mean ASRs are approximately
$98.3\%$, $97.0\%$, $96.9\%$, $95.5\%$, and $96.4\%$,
respectively. DFCS therefore retains at least $95.5\%$ mean ASR across
the five evaluated representation spaces, with DINOv3 providing the highest
mean and the smallest dispersion.

\subsubsection{Allocation Method}
Figure~\ref{fig:result-f} compares six allocation methods under the same experimental setting. Except for
$k$-center, their mean ASRs range from approximately $95.0\%$ to
$98.3\%$. The default $k$-means configuration achieves about $98.3\%$
ASR in $0.16$\,s. GMM is similarly effective but requires $2.3$\,s,
whereas BIRCH and agglomerative clustering require more than $120$\,s.

The results reflect differences in their allocation objectives. While
$k$-means targets average squared distortion over the candidate distribution,
$k$-center minimizes the worst-case covering radius and may allocate scarce
slots to low-density regions. Its approximately $51\%$ mean ASR
and high variability support average-distribution coverage as the
better-aligned objective in this low-budget setting. In contrast,
$k$-medoids restricts representatives to real samples during clustering,
coupling region formation with representative selection. Its lower mean ASR
suggests that this discrete coupling is less robust than DFCS's two-stage
design.

\subsection{Mechanism Diagnostics and Controlled Comparisons}

\paragraph{Purpose and protocol.}
This subsection tests whether DFCS exhibits the allocation properties
identified by Proposition~\ref{prop:coverage_main} and whether the corresponding
signal-error components vary consistently with the local decomposition. Under the
proposition's local assumptions and a common effect envelope, smaller
$\Delta_2$ or $\Gamma$ tightens the corresponding term in the upper bound on
equal-weight signal-approximation error. DFCS targets the coverage term through
its $k$-means and centroid-nearest construction, whereas $\Gamma$ is evaluated
as a complementary property of the selected set. Favorable values for both
terms therefore indicate a more faithful local equal-weight summary of the
candidate pool; end-to-end ASR is assessed separately through matched victim
training.

We instantiate these diagnostics on CIFAR-10/BadNets with target class 0 and
$B=20$. All diagnostics use the same non-target candidate pool, normalized
DINOv3 features, frozen clean ResNet-18 checkpoint, and 100 approximately
class-balanced non-target evaluation inputs. Local replacement effects are
computed in the final classifier-head parameter subspace. We record ten
configured subset runs for each of DFCS, the six main baselines, and four
structural controls, yielding 110 selector--run records.

\paragraph{Diagnostic quantities.}
For each subset, $\Delta_2$ measures the mean squared feature distance to
the nearest selected samples, while $\Gamma$ measures the mismatch
between represented candidate masses and uniform poison weights.
$E_{\mathrm{sig}}^{w}$ is the normalized error between the candidate-average and a region-mass-weighted representative signal, isolating
the coverage. $E_{\mathrm{mass}}$ is the normalized error between
the region-weighted and implemented equal-weight signals, isolating the
representative-mass component. They are normalized by the mean
absolute candidate-average signal. Their complete definitions are provided
in Appendix.

\paragraph{Structural controls.}
Class-balanced RS distributes the budget as evenly as possible across
eligible source classes. Histogram-matched RS randomly selects the same
number of samples from each source class as the DFCS subset.
Cluster-random retains DFCS's fitted $k$-means partition but samples one
candidate uniformly from each cluster instead of selecting the
centroid-nearest candidate. Class-stratified DFCS retains the same per-class
sample counts as DFCS but performs clustering separately within each
represented source class. These controls examine equal class allocation,
source-class composition, representative choice, and global versus
within-class coverage, respectively. DFCS attains the lowest mean geometric
diagnostics, while the matched victim comparisons generally support it over
these alternatives.

\paragraph{Component-wise results.}
Consistent with the proposition's component-wise consequence, we examine
descriptive selector-level associations between
$\sqrt{\overline{\Delta_2}}$ and $\overline{E}_{\mathrm{sig}}^{w}$ and between
$\overline{\Gamma}$ and $\overline{E}_{\mathrm{mass}}$.
Across the eleven selector means reported in
Appendix, the corresponding Spearman
correlations are $\rho=0.809$ and $\rho=0.445$. DFCS achieves the lowest mean
$\Delta_2$ and $\Gamma$ among all evaluated selectors and controls, reducing
them by $13.9\%$ and $28.5\%$, respectively, relative to histogram-matched
RS. Cluster-random also has higher mean values for both terms than DFCS.
These comparisons argue against source-class composition or the fitted
partition alone as complete explanations and are consistent with a benefit
from centroid-nearest representative selection.

\paragraph{Matched victim controls.}
The matched end-to-end comparisons also support DFCS. Under the five-runs, DFCS improves mean ASR over histogram-matched RS, cluster-random,
and class-stratified DFCS by $10.96\%$, $7.40\%$, and $1.39\%$, respectively, with paired wins in $5/5$, $5/5$, and $4/5$ runs.
Across DFCS and these controls, mean ACC ranges from $94.10\%$ to
$94.53\%$, close to the $94.38\%$ no-attack mean. Together, the
geometric diagnostics and matched victim comparisons support centroid-nearest
representative selection and a modest advantage for global over within-class
clustering in this setting. The diagnostics provide descriptive selector-level
evidence consistent with the local decomposition, while the victim runs
provide separate end-to-end evidence for attack behavior. Complete protocols,
formulas, and results are reported in
Appendix.

\section{Conclusion}

In this paper we introduce DFCS, which clusters frozen candidate features into $B$
clusters and selects the sample nearest each centroid, allocating equal-weight
dirty-label poisoning slots without task-specific selector training. Across
three datasets and two attacks, DFCS achieves the highest mean ASR in all
settings while preserving clean accuracy, with efficient sample selection. It
remains effective across the evaluated budgets, victim-training strategies,
targets, victim architectures, and encoders, while the local analysis and
matched controls supported global, centroid-nearest allocation. Our evaluation
is currently restricted to fixed, low-budget dirty-label image classification with
public pretrained encoders; future work would examine clean-label attacks and
broader trigger and defense settings.

\bibliography{aaai2027_review_revised}


\end{document}